\documentclass[twocolumn,showpacs,aps,epsfig]{revtex4}

\usepackage{graphicx}
\usepackage{epstopdf}
\usepackage{latexsym}
\usepackage{amssymb}

\usepackage[center]{subfigure}

\begin{document}

 \newcommand{\bq}{\begin{equation}}
 \newcommand{\eq}{\end{equation}}
 \newcommand{\bqn}{\begin{eqnarray}}
 \newcommand{\eqn}{\end{eqnarray}}
 \newcommand{\nb}{\nonumber}
 \newcommand{\lb}{\label}
\newcommand{\PRL}{Phys. Rev. Lett.}
\newcommand{\PL}{Phys. Lett.}
\newcommand{\PR}{Phys. Rev.}
\newcommand{\CQG}{Class. Quantum Grav.}

\title{On the motion of particles in Covariant Ho\v rava-Lifshitz Gravity and the meaning of the A-field} 

\author{Elcio Abdalla}
\email{eabdalla@fma.if.usp.br}
\author{Alan M. da Silva}
\email{amsilva@fma.if.usp.br}


\affiliation{Instituto de F\'isica, Universidade de S\~ao Paulo\\
C.P.66.318, CEP 05315-970, S\~ao Paulo, Brazil}

\date{\today}

\begin{abstract}

We studied the low energy motion of particles in the general covariant version of Horava-Lifshitz gravity proposed by Horava and Melby-Thompson. Using a scalar field coupled to gravity according to the minimal substitution recipe proposed by da Silva and taking the geometrical optics limit, we could write an effective relativistic metric for a general solution. As a result, we discovered that the equivalence principle is not in general recovered at low energies, unless the spatial Laplacian of A vanishes. Finally, we analyzed the motion on the spherical symmetric solution proposed by Ho\v rava and Melby-Thompson, where we could find its effective line element and compute spin-0 geodesics. Using standard methods we have shown that such an effective metric cannot reproduce Newton's gravity law even in the weak gravitational field approximation.

\end{abstract}

\maketitle

\section{Introduction}
In  \cite{Membranes, Horava1}, Ho\v rava proposed an anisotropic gravity theory inspired by the Lifshitz scalar \cite{lipstick} which has often been called Ho\v rava-Lifshitz (HL) gravity. In contrast with general relativity (GR), HL gravity has the advantage of being power-counting renormalizable.

HL theory is built on the basic assumption of anisotropic scaling between space and time, \emph{i.e.},
\begin{equation}
x^i \rightarrow b x^i, \qquad t \rightarrow b^{z} t
\label{scaling}
\end{equation}
\noindent
where $z$ is the dynamical critical exponent. The scaling dimension of an operator $\phi$ is defined by its transformation under (\ref{scaling}). If $\phi \rightarrow b^{-s}\phi$, then $[\phi]=s$ is the scaling dimension of $\phi$. We assume $z=D$, a necessary condition in order to achieve power counting in $(D + 1)$-dimensional gravity.

In a theory with anisotropic scaling, space and time are fundamentally distinct. It is thus natural to use the Arnowitt-Deser-Misner (ADM) formalism, splitting spacetime into space slices and time. In the ADM formalism the spacetime metric is decomposed as:
\begin{equation}
ds^2 = -N^2 dt^2 + \gamma_{ij}(dx^i +N^i dt)(dx^j + N^j dt).\label{metrica ADM}
\end{equation}
\noindent

Inspired by the ADM splitting, we build HL theory as a theory of the fields $N$, $N^i$ and $\gamma_{ij}$, which are, respectively, a scalar function, a 3-vector and a 3-dimensional metric tensor. The line element (\ref{metrica ADM}) built with those fields is not a fundamental object in HL theory, though, as we will show further in this paper, an effective line element arises in the low energy limit.

The spacetime anisotropy implies that GR's general diffeomorphism invariance does not fit HL theory naturally. Thus, we consider the local symmetries being restricted to
\begin{equation}
\delta t = f(t), \qquad \delta x^i = \xi^i(t, x^j), \label{DiffMF}
\end{equation}
\noindent
which are the foliation preserving diffeomorphisms, Diff$(\mathcal{M}, \mathcal{F})$, where $\mathcal{M}$ is the spacetime manifold, provided with a preferred foliation structure $\mathcal{F}$. 

If we restrict $N$ to depend only on time, $N = N(t)$ the theory is called projectable. Otherwise, for $N = N(t, x^i)$, we have the non-projectable HL theory. In this paper we just consider the
projectable version of HL theory.

The action of a $D+1$-dimensional HL theory has the form:
\begin{equation}
S=S_K - S_V,
\end{equation}    
\noindent
where  
\begin{eqnarray}
S_K &=& \int{dtd^D x\sqrt{\gamma} N \left[K_{ij}G^{ijkl}K^{kl}\right]}=\nonumber\\
&=&\int{dtd^D x\sqrt{\gamma} N \left[K_{ij} K^{ij} - \lambda K^2\right]}\label{cinetico}
\end{eqnarray} 
\noindent
is the kinetic term, which contains the time derivatives, with
\begin{equation}
G^{ijkl}=\frac{1}{2}\left( \gamma^{ik}\gamma^{jl}+\gamma^{il}\gamma^{jk}\right)-\lambda \gamma^{ij}\gamma^{kl}
\end{equation}
\noindent
being a generalized DeWitt metric. The $\lambda$ parameter comes from the absence of spacetime diffeomorphism symmetry due to anisotropic scaling, and its GR value is $\lambda =1$. A mechanism to make $\lambda$ be close to 1 in HL theory, at least at the IR limit, is necessary to match observational constraints\cite{dutta}, but such a mechanism is still unknown. Recently, it has been argued that $\lambda \neq 1$ may spoil the unitarity of quantum HL theory \cite{Bemfica}.

The spatial tensor $K_{ij}$ is the extrinsic curvature of spatial slices defined by:
\begin{equation}
K_{ij}=\frac{1}{2N}\left(\dot{\gamma}_{ij}-\nabla _i N_j - \nabla _j N_i \right)
\end{equation}
\noindent
where a dot indicates the time derivative, $\nabla$ is the covariant derivative on the spatial slice $\Sigma$, whose metric $\gamma_{ij}$ is used to raise and lower indices. 

The potential term $S_V$ is defined by:
\begin{equation}
S_V = \int{dtd^D x \sqrt{\gamma} N \mathcal{V}(\gamma_{ij})}
\end{equation}
\noindent
where $\mathcal{V}$ is a scalar operator built with the spatial metric and its spatial derivatives, with $[\mathcal{V}]=2z$. The most general $\mathcal{V}$ in projectable HL theory contains all operators with six spatial derivatives of the metric or less. The most general parity invariant potential is given in \cite{Sotiriou}.

\section{Covariant HL Gravity}

One of the issues of HL theory is the appearance of an extra degree of freedom, which has been called scalar graviton or spin-0 graviton. Although the scalar graviton may decouple in the $\lambda \rightarrow 1$ limit, as shown in \cite{MukohyamaWang} for projectable HL theory, it was the motivation for the construction of a version of HL theory with no extra degree of freedom, by Ho\v rava and Melby-Thompson \cite{Horava2}. This theory was originally called general covariant gravity at a Lifshitz point, but we will refer to it as the covariant HL theory, for short.

Ho\v rava and Melby-Thompson constructed the covariant HL theory for $\lambda=1$ adding a $U(1)$ extra gauge symmetry in the theory,
\begin{equation}
\delta_{\alpha} N = \delta_{\alpha} \gamma_{ij} =0, \qquad \delta_{\alpha} N_i = \nabla_i \alpha,
\label{alfatrans}
\end{equation}
\noindent
and introducting the gauge fields $A$ and $\nu$, which transform under $U(1)$ as:
\begin{equation}
\delta_{\alpha} A= \dot{\alpha} - N^i \partial_i \alpha, \qquad \delta_{\alpha} \nu= \alpha.
\end{equation}

The action of covariant HL theory for $\lambda=1$ is
\begin{eqnarray}
S[N,N^i, \gamma_{ij}, A, \nu]&=& \nonumber\\
 \int{dtd^Dx\sqrt{\gamma} \left[K^{ij}K_{ij}- K^2 + \mathcal{V}(\gamma_{ij})\right]} &+& S_{\nu}+S_{A},\nonumber\\
\end{eqnarray}
\noindent
where
\begin{eqnarray}
&S_{\nu}& = \int{dtd^D x \sqrt{\gamma}}N \times \nonumber\\
&\times& \nu \left(R^{ij}-\frac{1}{2}\gamma^{ij}R+ \Omega \gamma^{ij}\right)\left(2 K_{ij} + \nabla _i \nabla _j \nu\right),
\label{Snu}
\end{eqnarray}
\noindent
and
\begin{equation}
S_A=- \int{dt d^D x \sqrt{\gamma} A (R-2\Omega)},\label{SA}
\end{equation}
\noindent
with $R_{ij}$ being the 3-dimensional Ricci tensor related to $\gamma_{ij}$ and $R$ its trace.

\section{The Spherically Symmetric Solution}

Hereafter, we specialize to the case $D=z=3$. In \cite{JAlexandre1} and \cite{GSW}, the solutions with spherical symmetry in covariant HL gravity for the case $\lambda=1$ were found. Spherically symmetric solutions of covariant HL gravity coupled to electromagnetism are shown in \cite{BLW}. We will be interested in what was called type iii) solutions in \cite{JAlexandre1}. The IR limit of this class of solutions was deduced in \cite{Horava2} by a simple argument. We can chose a gauge using $U(1)$ symmetry that fixes $\nu =0$, eliminating $S_{\nu}$ from the total action. We obtain:
\begin{eqnarray}
&S&=\zeta^2\int{dtd^3x\sqrt{\gamma}N} \times \nonumber\\
&\times&[\left(K^{ij}K_{ij}-K^2 + R + 2\Lambda \right)-\frac{A}{N}\left(R-2\Omega \right)\nonumber\\
&+& O(\zeta^{-2})],
\end{eqnarray}
\noindent
where we have introduced the momentum scale $\zeta = (16 \pi G)^{-1/2} \sim 1/ l_P $ ($l_P$ denotes the Planck length) by making the redefinitions:
\begin{eqnarray}
t & \rightarrow & \zeta^{-2} t, \qquad N^i \rightarrow \zeta^{2}N^i \nonumber\\
A & \rightarrow & \zeta^4A, \qquad \alpha \rightarrow \zeta^{2} \alpha.
\end{eqnarray}

The IR limit corresponds to making $\zeta \rightarrow \infty$. The O($\zeta^{-2}$) terms in the action turn out to be negligible, eliminating all the high order spatial derivatives terms. If, in addition to the IR limit, we suppose $K_{ij}=0$ and $\Lambda= \Omega$, the action we obtain is:
\begin{equation}
S=\int{dtd^3x\sqrt{\gamma}\left(N-A\right)\left(R-2\Lambda\right)},\label{acaoestatica}
\end{equation}
\noindent
which is the GR action for $K_{ij}=0$, if we identify
\begin{equation}
\mathcal{N}(t,x^i) \equiv N(t)-A(t,x^i),
\end{equation}
where $\mathcal{N}$ is the general relativistic function lapse in the ADM formalism. In the case of vanishing cosmological constant, the Schwarzschild metric is a well-known solution of (\ref{acaoestatica}), with:
\begin{equation}
N=1, \quad A=1-\sqrt{f(r)}, \quad \gamma_{ij}dx^idx^j= \frac{dr^2}{f(r)}+r^2d\Omega^2,\label{solution}
\end{equation}
\noindent
where $f(r)=1-\frac{2M}{r}$. If we interpret solution (\ref{solution}) as equivalent to GR's Schwarzschild solution, this suggests that the gauge field $A(t,x^i)$ may be closely related to the gravitational potential. The way the $U(1)$ gauge theory can be understood as a limit of  diffeomorphisms of the type $\delta t = \xi(t,x^i)$, as shown in \cite{Horava2, JAlexandre1}, seems to reinforce this interpretation.

On the other hand, solution (\ref{solution}) has been interpreted by Greenwald \emph{et al.} in \cite{Greenwald} as a Einstein-Rosen bridge. According to this point of view, the A-field has no direct relation with the relativistic lapse, being simply a (time-vector and space-scalar) field  coupled with the ADM fields in the theory. However, they did not take into account the possible role of the coupling between matter and the A-field as we are doing here.

\section{Matter Coupling and Equivalence Principle}

In \cite{meupaper}, a recipe to build Lagrangians of fields coupled to covariant HL gravity was proposed. Using that recipe, the action of a scalar field takes the general form:
\begin{eqnarray}
&S&= \frac{1}{2} \times \nonumber\\
 &\times& \int{dtd^3x \sqrt{\gamma}N \left[\frac{(\dot{\phi}-\hat{N}^i\partial_i \phi )^2}{N^2} + c_0(\phi)\phi \Delta \phi + m^2 \phi^2 \right]} \nonumber\\
&+& \int{dtd^3x \sqrt{\gamma}\left[c_1(\phi)\Delta \phi + c_2(\phi)\nabla^i \phi \nabla_i \phi \right]\left(A-a\right)},\label{acaoescalar}
\end{eqnarray} 
\noindent
where $\Delta = \gamma_{ij}\nabla^i \nabla^j$,  $\hat{N}_i=N_i - \partial_i \nu$, $a = \dot{\nu}-N^j \nabla_j \nu + \frac{N}{2} \nabla^i \nu \nabla_i \nu$ and we omitted the high spatial derivative terms and a possible potential. We will be interested here in the weak field limit of this action, which amounts to consider only the terms in the action up to quadratic order in $\phi$. In this case we have $c_0(\phi) = c_0$, $c_1(\phi)= c_1 \phi$ and $c_2(\phi) = c_2$, where $c_0$, $c_1$ and $c_2$ are constants. To further simplify our equations, we choose the gauge $\nu=0$.

Under those assumptions, the $\phi$ equation of motion in the IR limit is
\begin{equation}
D^2 \phi + K D \phi - c_0 \Delta \phi + m^2 \phi - I(\phi, A)=0,\label{equacaoescalar}
\end{equation}
\noindent
where $D = \frac{1}{N} (\partial_t -N^i \partial_i)$ and
\begin{equation}
I(\phi, A) = b\left[ \frac{A}{N} \Delta \phi + \nabla_i \phi \nabla^i \left(\frac{A}{N} \right)\right] + c_1 \phi \Delta \left(\frac{A}{N} \right),\label{fieldequation}
\end{equation}
\noindent
with $b\equiv 2(c_1 - c_2)$.

Before the analysis of the special case (\ref{solution}), we consider the geometrical optics limit of the field $\phi$ in a general background, as a mean to find the equation of motion of classical particles in covariant HL theory. This procedure has already been discussed for projectable and non-projectable HL theory \cite{KK, Capasso, Suyama} and the results they have obtained are transfered for covariant HL theory only in case of vanishing $A$-field.   

To proceed with the geometrical optics approximation, we write
\begin{equation}
\phi(t,x^i) = R(t,x^i)e^{iS(t,x^i)},\label{ReiS}
\end{equation}
\noindent
insert (\ref{ReiS}) into (\ref{fieldequation}), consider the real part of it, suppose that $S$ derivatives are much larger than $R$ derivatives and keep only the leading order terms. We obtain:
\begin{eqnarray}
&-&\frac{1}{N^2}\left[ (\partial_t S)^2 - 2N^i \partial_i S \partial_t S + (N^i \partial_i S)^2 \right] \nonumber\\
&+& \left(c_0 +b \frac{A(t,x^i)}{N}\right)\gamma^{ij}\partial_i S \partial_j S  = \nonumber\\
&=&-m^2 + c_1 \Delta \left(\frac{A}{N} \right) + b \gamma^{ij} \frac{\partial_i R}{R}\partial_j \left(\frac{A}{N} \right).\label{equacaoS}
\end{eqnarray}

The last two terms on right-hand side of (\ref{equacaoS}) are not usual and must be addressed. The last one, $b \gamma^{ij} \frac{\partial_i R}{R}\partial_j \left(\frac{A}{N} \right)$, can be safely discarded as negligible as long as $\partial_j \left(\frac{A}{N} \right)$ is bounded in the considered spacetime region, as we suppose the limit $\partial R \ll \partial S$. The other one, $c_1 \Delta \left(\frac{A}{N} \right)$ cannot be discarded by the same argument and we must maintain it. 

We should remember that in the ADM formalism we have the following identities for the four-dimensional metric $g_{\mu \nu}$:
\begin{equation}
g^{00}= -\frac{1}{N^2}, \quad g^{0i}=\frac{N^i}{N^2}, \quad g^{ij}= h^{ij} - \frac{N^i N^j}{N^2}, \label{ADMsplit}
\end{equation}
\noindent
where $h_{ij}$ stands for the 3-dimensional metric tensor.

Equation (\ref{equacaoS}) is a Hamilton-Jacobi equation. Defining $x^{\mu}= (t,x^i)$, $p_{\mu}=\partial_{\mu}S$ and using (\ref{ADMsplit}), we can write:
\begin{equation}
\mathcal{H}(x^{\mu}, p_{\mu}, \tau)=-g^{\mu \nu}p_{\mu}p_{\nu}-m^2 + c_1\Delta\left(\frac{A}{N} \right)=0,\label{gmunu}
\end{equation} 
\noindent
where $g_{\mu \nu}dx^{\mu}dx^{\nu} \equiv ds^2_e$ is the effective line element:
\begin{equation}
ds^2_e=-N^2 dt^2 +\frac{\gamma_{ij}}{c_0+b\frac{A(x^{\mu})}{N}}\left(dx^i + N^i dt \right)\left(dx^j +N^jdt\right).\label{ds2e}
\end{equation} 

First, we must remark that (\ref{ds2e}) \emph{is different from the line element (\ref{metrica ADM}) from which we started}, unless $b A(x^{\mu})$ vanishes. This interpretation of (\ref{ds2e}) as an effective relativistic line element is consistent provided $c_0 +\frac{bA(x^{\mu})}{N} >0$.

Secondly, the super-Hamiltonian (\ref{gmunu}) contains a coordinate dependent potential given by $c_1 \Delta \left(\frac{A}{N} \right)$, where we remind that $\Delta$ is the Laplacian related to $\gamma_{ij}$ and not to the effective metric. It means that, in general, covariant HL gravity coupled with scalar matter does not respect the weak form of the principle of equivalence  \cite{Weinberg}, \emph{even in the IR limit}, as the equation of motion of otherwise free particles is mass dependent.  

The violation of the equivalence principle in the UV limit is a well-known property of HL-type theories, as the high energy corrections of the geodesic equation depend on $m/M_p$, where $M_p \sim \zeta$ is the Planck mass. Such corrections may be small enough at accessible energies to be compatible with experimental data. This is not the case for the IR violation of equivalence principle we found, which is a covariant HL only effect.

\section{Motion in the Schwarzschild-like background}

In this section we study the motion of particles in the background given by (\ref{solution}), using the method discussed above. We should first note that $\Delta A =0$ in this case, therefore the weak equivalence principle is satisfied and the motion of free particles will be mass independent in this background. Moreover, the coupling between $A$ and $\phi$ depends only on the parameter $b$.

Inserting the background (\ref{solution}) in (\ref{equacaoescalar}) we obtain:
\begin{eqnarray}
\partial_t^2 \phi &-& \left[ c_0  + b\left(1-\sqrt{f(r)}\right) \right]\Delta \phi \nonumber\\
&+& b f(r)\partial_r \sqrt{f(r)} \partial_r \phi  + m^2 \phi =0.\label{fieldequation1}
\end{eqnarray}

As we must obtain standard Klein-Gordon equation in flat spacetime in the limit $r \rightarrow \infty$, we set $c_0 = c^2 = 1$.

Using the substitution (\ref{ReiS}) and following the same steps, we find the particular case of equation (\ref{equacaoS}):
\begin{eqnarray}
 - (\partial_t S)^2  &+&\left[1 + b \left(1- \sqrt{f} \right) \right]  \nabla_i S \nabla^i S \nonumber\\
  &+& b f \partial_r \sqrt{f} \frac{\partial_r R}{R}  =-m^2.
\end{eqnarray}
The last term on left-hand side is not dangerous since $f \partial_r \sqrt{f} = \frac{1}{3} \partial_r f^{3/2}$ is positive for $r>2M$, bounded from above and behaves as $O(M/r^2)$ for large $r$. Thus, considering $\partial_r R$ small, we can safely discard this term as well and insert (\ref{solution}) into (\ref{ds2e}) to obtain the effective line element:
\begin{equation}
ds^2_{e}= -dt^2 + \left[1+b(1- \sqrt{f})\right]^{-1}\left(\frac{dr^2}{f} + r^2 d \Omega^2 \right),\label{metrica}
\end{equation}
\noindent
which, at first sight, has no resemblance with Schwarzschild spacetime.

To show that this is indeed the case, we must test this effective metric in well-known situations, using the standard GR tools. The cases of visible light or classical particle at low velocity in the Sun's exterior gravitational field both fit our approximations very well and could, at least in principle, lead us to restrain the value of $b$, which is the only free parameter in our equations.

After a standard calculation, we obtain, at first order, the deflection of massless particles in the geometry given by (\ref{metrica}) as
\begin{equation}
\delta \phi = \frac{4M}{r_0}\left(\frac{2-b}{4}\right),
\end{equation}
\noindent
where $r_0$ is the impact parameter. This coincides with the GR result for $b=-2$. This coincidence, despite the differences between (\ref{metrica}) and Schwarzschild spacetime, is due to the fact that null geodesics are invariant by conformal transformations and (\ref{metrica}) is conformally related to
\begin{equation}
ds_c^2 = - \left(1+b-b\sqrt{f}\right)dt^2+ \frac{dr^2}{f}+r^2d\Omega^2,
\end{equation}
\noindent
which can be expanded for small $M/r$ as:
\begin{equation}
ds_c^2= - \left[1+\frac{bM}{r} + O\left(\frac{M^2}{r^2}\right)\right]dt^2 + \frac{dr^2}{f} + r^2 d\Omega^2,
\end{equation}
\noindent
which coincides with Schwarzschild metric at leading order for $b=-2$.

If we use massive test particles the result is different, as conformally related spacetimes are inequivalent. Considering, for instance, a radial motion of a massive particle in (\ref{metrica}), we have
\begin{equation}
-\dot{t}^2 + \left[ 1+b(1-\sqrt{f}) \right]^{-1}f^{-1} \dot{r}^2 = -1,\label{equacaoradial}
\end{equation}
\noindent
where the dot stands for $d/d \tau$ and $\tau$ is the proper time. The conserved quantity related to time independence of (\ref{metrica}) is $\dot{t} \equiv \mathcal{E}$. Substituting into (\ref{equacaoradial}), we obtain
\begin{equation}
\frac{1}{\left[1+b(1-\sqrt{f})\right]f} \dot{r}^2 = \mathcal{E}^2 - 1. \label{patologica}
\end{equation}

Equation (\ref{patologica}) is clearly incompatible to what we know about gravitational physics. Consider, for example, a particle with initial conditions $r(\tau=0)=r_0$ and $\dot{r}(\tau=0) = 0$, that is, a particle dropped at a point $r_0$. According to (\ref{patologica}) this particle will stay indefinitely at rest!

It is instructive to write equation (\ref{patologica}) as sum of kinetic and potential energy, to first order in $M/r$:
\begin{equation}
\frac{1}{2}\dot{r}^2 + E \left(2-b \right)\frac{M}{r}=E,\label{patologica2}
\end{equation}
\noindent
where $E=\frac{1}{2}(\mathcal{E}^2 - 1 )$ is the mechanical energy per unit mass. We can interpret (\ref{patologica2}) as an energy dependent gravitational constant, and we have $E=0$ in the case of a dropped particle, hence, a particle at rest does not feel any gravitational field, thus stays still . If we interpret (\ref{solution}) as the gravitational field of a central mass, this is utterly irreconcilable with Newton's gravitational law, which should been valid at the limit of weak gravitational field and low velocities, as we had taken. It is worth noting also that the force can be repulsive as well, depending on the value of $b$.

\section{Final Remarks}

We conclude our analysis by stating that the interpretation of the  $A$-field given in \cite{Horava2}, as a part of the GR function lapse, $\mathcal{N}=N-A$ is not compatible with the prescription for coupling covariant HL gravity to matter proposed by one of the authors in \cite{meupaper}. If we do not stick with this interpretation of $A$, we need not interpret (\ref{solution}) as a Schwarzschild-like spacetime in HL gravity. However, it is still a spherically symmetric solution of the theory and if covariant HL gravity effectively describes our universe, we
should study the stability of this solution to know whether it is or it is not expected to be found in nature. We must remind the reader that (\ref{patologica}) is not the only spherically symmetric solution of covariant HL theory. One of the solutions found in \cite{JAlexandre1, GSW} is just Schwarzschild solution in Painleve-Gullstrand coordinates, with $A=\nu=0$, whose effective IR line element is given by:
\begin{equation}
ds^2 = -dt^2 + (dr + N^r dt)^2 +r^2(d\theta^2 + \sin^2 \theta d \varphi^2),
\end{equation}
\noindent
with $N^r = \sqrt{2M/r}$. In this case, at low energy, our scalar field propagates in the same way it does in the relativistic case, leading to standard (and mass independent) Schwarzschild geodesics in the geometrical optics limit.

Instead, it may be that the coupling recipe we used is inadequate and we should find another way to couple matter to gravity in covariant HL theory. The fact that we do not recover the weak equivalence principle in general seems to be an issue of such a recipe. Maybe we should look for a way to relate U(1) transformations to boosts in Lorentz group, as the one suggested by Ho\v rava and Melby-Thompson in \cite{Horava2}. The Minkowski solution in covariant HL gravity is:
\begin{equation}
N=1, \quad N^i =0,\quad g_{ij}= \delta_{ij},\quad A = \nu =0.\label{plano}
\end{equation}

Since standard Lorentz boosts do not belong to Diff($\mathcal{M},\mathcal{F}$), they proposed that a Lorentz boost in covariant HL theory should be composed of a foliation preserving diffeomorphism and a U(1) transformation. Thus, a boost-like transformation in the x-direction should be written:
\begin{eqnarray}
t^{\prime}&=& t\cosh \omega , \nonumber\\
x^{\prime}&=& x\cosh \omega  - t\sinh \omega , \nonumber\\
\alpha &=& -x\sinh \omega. \label{boost}
\end{eqnarray}

Solution (\ref{plano}) is invariant under transformation (\ref{boost}) with the exception of the field $\nu$, which defines a preferred frame. The standard action of a scalar field in flat spacetime,
\begin{equation}
S[\phi]= \frac{1}{2}\int{dtd^3x \left[(\partial_t \phi )^2 - ( \vec{\partial} \phi)^2\right]},
\end{equation}
\noindent
is invariant under (\ref{boost}), as it contains no coupling with $\nu$. A coupling between $\phi$ and the invariant $A-a$ (which vanishes in (\ref{plano})) may appear in the general case, as in the recipe used to build (\ref{acaoescalar}). On the other hand, it seems that in this case matter should couple to $N_i$ instead of the invariant $\hat{N}_i = N_i - \partial_i \nu$ we have used here. However, it is not clear how we could build a full U(1)$\times$Diff($\mathcal{M},\mathcal{F}$) invariant action of a scalar coupled to a general solution in this manner.

Those two questions, the meaning of the A-field and the coupling of covariant HL gravity to matter, are still widely open and deserve further investigation.

\noindent
\textbf{Acknowlegments:} This work was partially supported by FAPESP and CNPq.


\end{document}